%% file: MWG2011-3_mod.tex
\documentclass[twocolumn]{pasj00}
\Received{2011 October 21}
\Accepted{2012 July 15}
\Published{$\langle$publication date$\rangle$}
\SetRunningHead{Honma et al.}{Fundamental Parameters of the Galaxy}

\usepackage{times}

\begin{document}

\title{Fundamental Parameters of the Milky Way Galaxy Based on VLBI astrometry}
\author{
Mareki \textsc{Honma},\altaffilmark{1,2}
Takumi \textsc{Nagayama},\altaffilmark{1}
Kazuma \textsc{Ando},\altaffilmark{3}
Takeshi \textsc{Bushimata},\altaffilmark{1}
Yoon Kyung \textsc{Choi},\altaffilmark{4}
Toshihiro \textsc{Handa},\altaffilmark{3}\\
Tomoya \textsc{Hirota},\altaffilmark{1,2}
Hiroshi \textsc{Imai},\altaffilmark{3}
Takaaki \textsc{Jike},\altaffilmark{1,2}
Mi Kyoung \textsc{Kim},\altaffilmark{1}
Osamu \textsc{Kameya},\altaffilmark{1,2}
Noriyuki \textsc{Kawaguchi},\altaffilmark{1,2}\\
Hideyuki \textsc{Kobayashi},\altaffilmark{1,2,5}
Tomoharu \textsc{Kurayama},\altaffilmark{3,6}
Seisuke \textsc{Kuji},\altaffilmark{1}
Naoko \textsc{Matsumoto},\altaffilmark{1}
Seiji \textsc{Manabe},\altaffilmark{1,2}\\
Takeshi \textsc{Miyaji},\altaffilmark{1}
Kazuhito \textsc{Motogi},\altaffilmark{7}
Akiharu \textsc{Nakagawa},\altaffilmark{3}
Hiroyuki \textsc{Nakanishi},\altaffilmark{3}
Kotaro \textsc{Niinuma},\altaffilmark{1,8}\\
Chung Sik \textsc{Oh},\altaffilmark{9}
Toshihiro \textsc{Omodaka},\altaffilmark{3}
Tomoaki \textsc{Oyama},\altaffilmark{1}
Nobuyuki \textsc{Sakai},\altaffilmark{1,2}
Katsuhisa \textsc{Sato},\altaffilmark{1}
Mayumi \textsc{Sato},\altaffilmark{4}\\
Katsunori M. \textsc{Shibata},\altaffilmark{1,2}
Satoshi \textsc{Shiozaki},\altaffilmark{3}
Kazuyoshi \textsc{Sunada},\altaffilmark{1}
Yoshiaki \textsc{Tamura},\altaffilmark{1,2}
Yuji \textsc{Ueno}\altaffilmark{1}
and Aya \textsc{Yamauchi}\altaffilmark{1}
}
\altaffiltext{1}{Mizusawa VLBI Observatory, NAOJ, Mizusawa, Ohshu, 023-0861}
\altaffiltext{2}{Department of Astronomical Science, Graduate University for Advanced Studies, Mitaka, 181-8588}
\altaffiltext{3}{Graduate School of Science and Engineering, Kagoshima University, Korimoto, Kagoshima 890-0065}
\altaffiltext{4}{Max-Planck-Institute for Radio Astronomy, Auf dem
H\"{u}gel 69, D-53121, Bonn, Germany}
\altaffiltext{5}{Department of Astronomy, The University of Tokyo, Hongo, Bunkyo, Tokyo 113-0033}
\altaffiltext{6}{Center for Fundamental Education, Teikyo University of Science,
Uenohara, Yamanashi, 409-0193}
\altaffiltext{7}{Graduate School of Science, Hokkaido University, N10 W8, Sapporo 060-0810}
\altaffiltext{8}{Department of Physics, Yamaguchi University, Yoshida, Yamaguchi, Yamaguchi 753-8512}
\altaffiltext{9}{Korean VLBI Network, KASI, Seoul, 120-749, Republic of Korea}


\email{mareki.honma@nao.ac.jp}

\maketitle

\begin{abstract}
We present analyses to determine the fundamental parameters of the
 Galaxy based on VLBI astrometry of 52 Galactic maser
 sources obtained with VERA, VLBA and EVN.
We model the Galaxy's structure with a set of parameters including the
 Galaxy center distance $R_0$, the angular rotation velocity at the LSR
 $\Omega_0$, mean peculiar motion of the sources with respect to
 Galactic rotation ($U_{\rm src}$,
 $V_{\rm src}$, $W_{\rm src}$), rotation-curve shape index, and the
 V component of the Solar peculiar motions $V_\odot$.
Based on a Markov chain Monte Carlo method, 
we find that the Galaxy center distance is constrained at a 5\% level to
 be $R_0=8.05 \pm 0.45$ kpc, where the error bar includes both
 statistical and systematic errors.
We also find that the two components of the source peculiar motion
 $U_{\rm src}$ and $W_{\rm src}$ are fairly small compared to the
 Galactic rotation velocity, being $U_{\rm src}=1.0 \pm 1.5$ km~s
 $^{-1}$ and $W_{\rm src}=-1.4\pm 1.2$ km~s$^{-1}$.
Also, the rotation curve shape is found to be basically flat
 between Galacto-centric radii of 4 and 13 kpc.
On the other hand, we find a linear relation between $V_{\rm src}$ and
 $V_\odot$ as  
$V_{\rm src}=V_\odot-19 \; (\pm2)$ km~s$^{-1}$, suggesting that the
value of $V_{\rm src}$ is fully dependent on the adopted value of $V_\odot$.
Regarding the rotation speed in the vicinity of the Sun, we also find
 a strong correlation between $\Omega_0$ and $V_\odot$.
We find that the angular velocity of the Sun, $\Omega_{\odot}$, which is
 defined as $\Omega_\odot \equiv \Omega_0 + V_\odot/R_0$, can be well
 constrained with the best estimate of $\Omega_\odot=31.09 \pm 0.78$ km
 s$^{-1}$ kpc$^{-1}$. This corresponds to $\Theta_0 = 238 \pm 14$ km s$^{-1}$ if
 one adopts the above value of $R_0$ and recent determination of
 $V_\odot$ $\sim$12 km s$^{-1}$.
\end{abstract}

\section{Introduction}

As a first-order approximation, the Milky Way Galaxy can be regarded as
an axi-symmetric system with a circular rotation, and its structure and
motions can be described combinations of fundamental parameters.
For instance, historically the Milky Way's rotation was discovered based
on the determinations of so-called Oort constants $A$ and $B$, which are
combinations of the distance to the Galaxy center $R_0$, and the rotation
velocity of the LSR (Local Standard of Rest) $\Theta_0$, and their derivatives.
Today the Galactic constants $R_0$ and $\Theta_0$ are most
frequently-used parameters which provide fundamental scales of the
Galaxy's size and rotation speed.
The IAU recommendation values are 8.5 kpc and 220 km~s$^{-1}$,
(Kerr \& Lynden-Bell 1986), but accurate values of these parameters are
still at issue: for instance, most of recent studies of $R_0$ favor
values somewhat smaller than the IAU value (Reid 1993: Ghez et al. 2008;
Gilssen et al. 2009; McMillan \& Binney 2010), being
between 8.0 and 8.5 kpc.
The rotation curve is another important quantity to describe the
rotation and mass distributions in the Galaxy, and often parameterized
with a few parameters (e.g., flat rotation curve or power-law rotation
curve).
However, the exact shape of the rotation curve also remains yet to be studied.

Astrometry of Galactic sources is believed to be most powerful to
determine the fundamental parameters of the Galaxy because it provides full
six-dimensional phase-space information of Galactic objects.
Most notably, until recently most of parallax measurements 
have been limited to local
sources, within $\sim$1 kpc from the Sun (for instance, such as using
HIPPARCOS).
However, recent development of high-accurate astrometry with VLBI is
now providing tens of sources for which precise distances and proper
motions are known.
Thanks to this, new determination of fundamental parameters
of the Milky Way Galaxy becomes feasible.
For instance, Reid et al. (2009b) used 18 maser sources to explore the
structure of the Milky Way and obtained the Galactic parameters such as
$R_0=8.4\pm0.6$ kpc and $\Theta_0=254\pm16$ km~s$^{-1}$.
In addition, they suggested that rotation of star-forming regions could
lag behind the true Galactic rotation by $\sim$15 km~s$^{-1}$.
There are also several studies of Galactic structures based on 
reanalysis of Reid et al.' data (e.g., Bobylev \& Bajkova 2010;
McMillan \& Binney 2010), which emphasize importance of Galactic
scale astrometry.
Since VLBI astrometry has been now extensively carried out with VERA
and VLBA, the number of sources for which astrometric information is
available is significantly increased (e.g., there are $\sim$10 new sources
reported in the VERA special issue in PASJ 63, 2011).
Therefore, it is worth conducting a new study of fundamental parameters
of the Galaxy based on these new results.
To carry out this, in this paper we compile the most-updated list of VLBI
astrometry results, and provide new analyses of fundamental parameters
of the Milky Way Galaxy.

\section{Sample}

\begin{figure*}[t]
\begin{center}
       \FigureFile(180mm,180mm){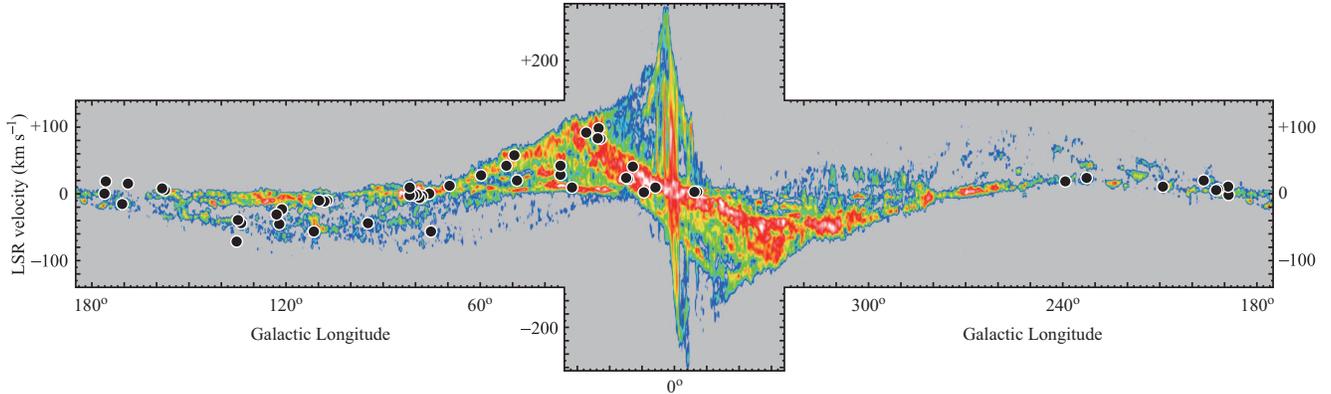}
\end{center}
\caption{Location of 52 maser sources for which accurate
 astrometric data are available (table 1), superposed on
 the longitude-velocity diagram of CO (Dame et al. 2001).}
\end{figure*}

We collected results of VLBI astrometry from the literatures
including most recent results, and compiled 
a list of parallaxes and proper motions in table 1.
The list consists of star-forming regions with maser sources or
continuum sources (such as T-Tau stars) that have been observed with
VERA (VLBI Exploration of Radio Astrometry)/VLBA
 (Very Long Baseline Array)/EVN (European VLBI Network).
The star-forming regions included here are those with H$_2$O and/or
CH$_3$OH maser emissions, or gyro-synchrotron emissions which originate
from magnetic activity of pre-main sequence stars.
We include in our sample almost all published works
available to date: 
the only exceptions are: 1) Sgr B2 (Reid et al.2009c), which is too close to
the Galaxy center (expected to be at $\sim$100 pc) and is known to have
significant peculiar motion, making this sources unsuitable for the
study of the Galactic rotation,
2) G34.43+0.24 (Kurayama et al. 2011) for which the proper motion in
declination is not measured due to poor UV coverage for this nearly-0
declination source,
 and 3) IRAS 05137+3919 (Honma et
al. 2011), which is likely to be located in the far outer
Galaxy ($R\ge 15$ kpc), where our simple parameterization of Galactic
rotation (e.g., power-low rotation curve) may not be appropriate.

We only selected star-forming regions, and excluded late-type stars
because properties of these populations are quite different from
those of star-forming regions: for instance, late-type stars are known
to have a larger velocity dispersion than that of star-forming regions, and 
their rotation lags behind the Galactic rotation, which is
known as {``}asymmetric drift{''}.
However, we note that we included two super-giant sources VY CMa (Choi
et al. 2008) and S Per (Asaki et al. 2010) since they are massive
stars that evolve quickly after their birth, and thus can be considered as
{``}young{''} populations.

Table 1 summarizes the 52 sources that are selected in the
manner described above.
In order to show the distribution of the sample sources in the
Galaxy, in figure 1 we show maser positions in the CO $l$-$v$
diagram obtained by Dame et al.(2001).
As can be seen in figure 1, the sources are mostly located in the
northern part of the Galactic plane.
This is because the observing arrays (VERA, VLBA and EVN) are all located
in the northern hemisphere.
However, except for the concentration to the northern hemisphere, the
sources are scattered well from the Galaxy center to the region beyond the
anti-center, up to $l\sim 240^\circ$.
To fill up the gap in the southern hemisphere, astrometry of
maser sources with LBA (Australian VLBI array) and forthcoming SKA 
(Square Kilometer Array)
will be effective, and this will be one of the most important future
tasks in this kind of study.

As we will see in later sections, the whole set of 52 sources may include
{``}outliers,{''} which have large deviations from expected motions.
Such a source could introduce systematic error into our results, and
thus its effect should be investigated carefully.
To do this, in later section (section 5), we will present results of 
various sample sets, in which some sources in list 1 are discarded
depending on its degree of deviations.
The source elimination will be made based on source's contribution to
the likelihood, $P_i(\mathbf{O}_i|\mathbf{M})$ (see equation [3]):
we create a new set of sample with $N-1$ sources by removing the source
with smallest $P_i(\mathbf{O}_i|\mathbf{M})$ in the $N$-source sample.
In this way, we will try various sets of samples with source numbers
ranging from 52 to 44 (for details see section 5).
In table 1 we mark the sources with their ranks in the likelihood.

\section{Galaxy model}

In order to derive fundamental Galaxy parameters, we use a simple
Galaxy model in which motions of sources are basically in 
circular rotation with small systematic/random motions.
As is widely used, the Local Standard of Rest (LSR) is set to be 
in circular rotation around the Galaxy center, which is described with
two Galactic constants $R_0$ and $\Theta_0$.
Obviously $R_0$ is the distance between the Galaxy center and the LSR, and
$\Theta_0$ is the Galactic rotation velocity at the LSR.
The ratio of $\Theta_0$ to $R_0$ gives the Galactic angular velocity at
the LSR $\Omega_0$ (=$\Theta_0/R_0$).
In the following analyses, we use a set of $R_0$
and $\Omega_0$, rather than a set of $R_0$ and $\Theta_0$, as the
Galactic constants to be solved because it is known that $R_0$ and $\Theta_0$
are tightly correlated (e.g., Reid et al.2009b; McMillan and Binney 2010).

Regarding the Galactic rotation curve, we adopt two different rotation
curves, a power law model and 2nd-order polynomial model.
In the power law model (which is mainly studied here),
the Galactic rotation velocity at any radius $R$ is assumed to obey the
following rotation low,
\begin{equation}
 \Theta(R) = \Theta_0 \left(\frac{R}{R_0}\right)^\alpha.
\end{equation}
Here $\alpha$ is the rotation curve index, and this parameter is 
to be solved in the following analysis to determine the shape of
the rotation curve.
Obviously, a flat rotation curve with $\Theta(R)=\Theta_0$ corresponds to
$\alpha=0$.
Note that $R$ is the radius in the cylindrical coordinate
system defined with (R, $\beta$, Z), with its origin located at the
Galaxy center (the notations here follow those in Reid et al. 2009b).
On the other hand, in order to assess the effect of rotation curve models
on Galactic parameter determinations, we also use the polynomial model
described as,
\begin{equation}
\label{eq:2pol-RC}
 \Theta(R) = \Theta_0 + a_0 (R-R_0) + b_0 (R-R_0)^2.
\end{equation}
Here $a_0=d\Theta/dR$ and $b_0=(d^2\Theta/dR^2)/2$, respectively, and
both coefficients are defined at $R=R_0$.
This rotation curve contains two parameters while the power law model
has only one.
Thus, the polynomial rotation curve has a potential to give a better fit
to the data, and we can test if inclusion of an additional parameter
would alter the results based on the power law model.

We also include possible systematic motions of star-forming regions as
model parameters to be solved, since Reid et al.(2009b) reported that
high mass star-forming regions would lag behind Galactic rotation by
$\sim 15$ km~s$^{-1}$.
In the present paper, this effect is re-considered in the same manner
as Reid et al.(2009b) by including
mean systematic motions of star-forming regions ($U_{\rm src}$,
$V_{\rm src}$, $W_{\rm src}$).
Note that the directions of $U_{\rm src}$, $V_{\rm src}$, $W_{\rm src}$
are defined at each source position, with $U_{\rm src}$ toward the
Galaxy center, $V_{\rm src}$ toward the direction of Galactic rotation, and 
$W_{\rm src}$ toward the north Galactic pole, respectively.

In addition to the above parameters that describe the motions of maser
sources in the Galaxy, we add one additional parameter to be solved for
in our analysis:
the $V$ component of the Solar motion with respect to the LSR ($V_\odot$).
As is widely known, the Sun is not at rest in the LSR frame but has
its own peculiar motion with respect to the LSR.
Based on the HIPPARCOS observations of local stellar motions and distances,
Dehnen and Binney (1998) obtained the Solar motion with respect to the 
LSR as, ($U_\odot$, $V_\odot$, $W_\odot$) = (10.00$\pm$0.36,
5.25$\pm0.62$, 7.17$\pm$0.38) km~s$^{-1}$, reducing the $V$ component of
the Solar motion by $\sim 10$ km~s$^{-1}$ from the classical value of
$V_\odot$=15.4 km~s$^{-1}$ (Kerr \& Lynden-Bell 1986).
However, recent reanalyses of the data reported that the $V$ component
of the Solar motion by Dehnen \& Binney (1998) was an underestimate and 
that an upward modification of $V_\odot$ is required by $\sim 7$ km~s$^{-1}$
(Sch\"{o}nrich et al. 2010), yielding $V_\odot\sim 12$  km~s$^{-1}$.
Interestingly, this upward modification reduces the rotation velocity
lag of the star-forming regions (McMillan \& Binney 2010).
Therefore, the lag parameter $V_{\rm src}$ and $V_\odot$ are correlated
to each other to some extent, and we include $V_\odot$ as possible 
parameter (fixed or solved depending on models considered) to see how
our results are affected.
Note that the other two components of the Solar motion ($U_\odot$,
$W_\odot$) are assumed to be
well-determined, since previous studies were consistent between
each other (e.g., Kerr \& Lynden-Bell 1986; Dehnen \& Binney 1998;
Aumer \& Binney 2009; Sch\"{o}nrich et al. 2010).
Throughout the present paper, we simply take $U_{\rm sun}$ and 
$W_{\rm sun}$ of Dehnen \& Binney (1998).
Table 2 summarizes the parameters to be solved for in the present analyses.

\setcounter{table}{1}
\begin{table*}
\begin{center}
\caption{Summary of the Galactic parameters used in the present paper.}
\begin{tabular}{ll}
\hline\hline
parameter & notes \\ 
\hline
$R_0$ & the distance from the Sun to the Galaxy center \\
$\Omega_0$ & Angular rotation velocity at the LSR ($\Omega_0=\Theta_0/R_0$)\\
$\alpha$ & rotation curve index defined as $\Theta(R)=\Theta_0(R/R_0)^\alpha$ \\
$a_0$, $b_0$ & rotation curve coefficients defined as
 $\Theta(R)=\Theta_0 + a_0(R-R_0) +b_0(R-R_0)^2$ \\
$U_{\rm src}$ & mean peculiar motion of SFRs toward the Galaxy
 Center \\
$V_{\rm src}$ & mean peculiar motion of SFRs toward the direction of
 Galactic rotation \\
$W_{\rm src}$ & mean peculiar motion of SFRs toward the north
 Galactic pole \\
$V_\odot$ & $V$ component of the Solar motion (toward the Galactic rotation)\\
\hline
\end{tabular}
\end{center}
\end{table*}

\section{Analysis}

To obtain the best estimates of the Galactic parameters described above, we
utilize a Markov chain Monte Carlo (MCMC) method with the Metropolis
algorithm.
To do this, first we define the likelihood function as
\begin{equation}
\label{eq:likelihood}
 L(\mathbf{M})=\prod_i P_i(\mathbf{O}_i|\mathbf{M}).
\end{equation}
Here $P_i(\mathbf{O}_i|\mathbf{M})$ is the conditional probability of finding
observables $\mathbf{O}_i$ given a set of model parameters $\mathbf{M}$.
$\mathbf{O}_i$ represents the observables for the $i$-th
source, a detailed description of which is provided later.
The model $\mathbf{M}$ represent the model parameters described in the
previous section: for the
power law rotation curve,
\begin{equation}
 \mathbf{M}=(R_0, \Omega_0, \alpha, U_{\rm src}, V_{\rm src}({\rm or}\,
  V_\odot), W_{\rm src}),
\end{equation}
or for the polynomial rotation curve case,
\begin{equation}
 \mathbf{M}=(R_0, \Omega_0, a_0, b_0, U_{\rm src}, V_{\rm src}({\rm or}\,
  V_\odot), W_{\rm  src}).
\end{equation}
Note that as we will see later there is a strong correlation between
$V_{\rm src}$ and $V_\odot$, and thus these two parameters are not
solved for at the same time, but one of them are fixed while the other
is being solved for.

In a practical MCMC method,  we calculate the likelihood $L(\mathbf{M})$
based on the equation (\ref{eq:likelihood}) for a current set of model
parameters $\mathbf{M}$.
Then, the next set of $\mathbf{M}'$ is calculated by adding a small
random fluctuations to the previous $\mathbf{M}$, and the likelihood
$L'(\mathbf{M}')$ is calculated.
If $L'(\mathbf{M}') \ge L(\mathbf{M})$, the next parameter $\mathbf{M}'$
is accepted. 
If not, we draw a random variable $u$, which has a uniform probability
between 0 and 1, and we accept the next parameter $\mathbf{M}'$ in case
of $L'(\mathbf{M}')/ L(\mathbf{M}) > u$.
In other case (i.e.,  $L'(\mathbf{M}')/ L(\mathbf{M}) \le u$), the next
parameter set is rejected and the parameter set remains at the previous
one $\mathbf{M}$.
Above procedures are iterated for a large number of trials (typically 
for $\sim 10^5$).
The best estimates of $\mathbf{M}$ and their error bars
are obtained by taking means and standard deviations of the samples of 
$\mathbf{M}$ in the MCMC trials, after removing early trials in the
initial {``}burn-in{''} phase, which is set to be the first 10\% of the
trials. 
Note that these procedures correspond to a MCMC method with a uniform
{\it a priori} probability distribution of model parameters, namely
$P(\mathbf{M})=const.$
The reason for adopting a constant prior probability is that in the
present paper we want to constrain Galactic parameters based only on
VLBI astrometry data of star-forming regions rather than partially
relying on other results in previous studies.
However, since the parameters $\mathbf{M}$ are well constrained
to relatively narrow ranges in the final results of MCMC (see next section)
the choice of a different {\it a priori} probability
(such as uniform probability in the logarithm) would not
alter the main results presented here.

For the observables, $\mathbf{O}_i$ (which is for the $i$-th source) is
given as
\begin{equation}
\label{eq:observables}
 \mathbf{O_i}=(V_{{\rm LSR}},\, \mu_{l},\, \mu_{b})_i.
\end{equation}
Here $V_{\rm LSR}$ is the radial velocity with respect to the LSR
(based on the conventional Solar motion), $\mu_l$ and $\mu_b$ are
the two components of proper motions along the Galactic coordinate $l$ and
$b$.
We adopted these quantities as observables because they are commonly used,
but here we note that they are not in the exactly same frame: 
while $V_{\rm LSR}$ is measured in the conventional LSR frame, $\mu_l$
and $\mu_b$ are observed in the heliocentric frame.
When MCMC analyses are done in later section, of course $\mu_l$,
$\mu_b$ and $V_{\rm LSR}$ are converted to the appropriate LSR frame by
using the adopted Solar motion.
We also note that VLBI astrometry provides the full 6-dimensional
coordinates in the phase space, but here the source coordinates $l$ and $b$
are not considered as {''}observables{''}.

In order to calculate the conditional probability
$P_i(\mathbf{O}_i|\mathbf{M})$ in equation (\ref{eq:likelihood}), 
here we simply adopt Gaussian
for the probability distribution of finding an observable given the model
$\mathbf{M}$ and the source parallax $\pi_i$, namely,
\begin{equation}
\label{eq:pdf}
 P(x_i|\mathbf{M},\pi_i)=\frac{1}{\sqrt{2\pi\sigma^2}}
  \exp \left({\frac{-(x_{\rm obs} -  x_{\rm model})^2}{2\sigma^2}}\right),
\end{equation}
where $x_{\rm obs}$ is the observable (corresponding to $V_{{\rm LSR}}$,
$\mu_{l}$, and $\mu_{b}$ of the $i$-th source).
The model value $x_{\rm model}=x_{\rm model}(\mathbf{M},\pi_i)$ is
calculated based on the Galactic
model $\mathbf{M}$ as well as the observed source parallax $\pi_i$.
The conditional probablity $P_i(\mathbf{O}_i|\mathbf{M})$ in equation
(\ref{eq:likelihood}) can be obtained
by multiplying the Gaussian distribution in equation (\ref{eq:pdf}) 
for the three parameters 
($V_{{\rm LSR}}$, $\mu_{l}$, and $\mu_{b}$) and also by integrating over
the possible parallax range of $\pi$.
The integration over the possible range of $\pi$ is  
done by adopting Gaussian distribution of $\pi$ with a 
standard deviation given by the observational uncertainty listed in
table 1.
Then, the likelihood in equation (\ref{eq:likelihood}) is evaluated by
multiplying the conditional probability $P_i(\mathbf{O}_i|\mathbf{M})$
for all the sources in the sample.

The standard deviation $\sigma$ in equation (\ref{eq:pdf}) is
determined by considering
both the errors in the observations and the uncertainties in model estimates.
In addition
to the observational errors listed in table 1, we include the effect of
source random motion, which is assumed to be isotropic with a
1-dimensional velocity dispersion of $\sigma_v$.
This velocity dispersion accounts for the turbulent motion in the
molecular clouds and also systematic difference of maser source motions from
proto-stars, and throughout the paper we set $\sigma_v=7$ km~s$^{-1}$
unless otherwise noted.
As for radial velocities, observational error of the LSR velocity
$\sigma_{\rm LSR, obs}$ and the source velocity
dispersion $\sigma_v$ is summed in quadrature to calculate $\sigma$ in
equation (\ref{eq:pdf}) as $\sigma_{\rm LSR}^2=\sigma_{\rm LSR,
obs}^2 + \sigma_v^2$. 
For proper motions, again we add the observational error $\sigma_{\rm
obs}$ and the random motions in quadrature to obtain $\sigma_{\mu\; l,b} =
\sqrt{\sigma_{{\rm obs\;} l,b}^2+(\sigma_v/D)^2}$.

\section{Results}

\begin{figure*}[h]
\begin{center}
       \FigureFile(125mm,125mm){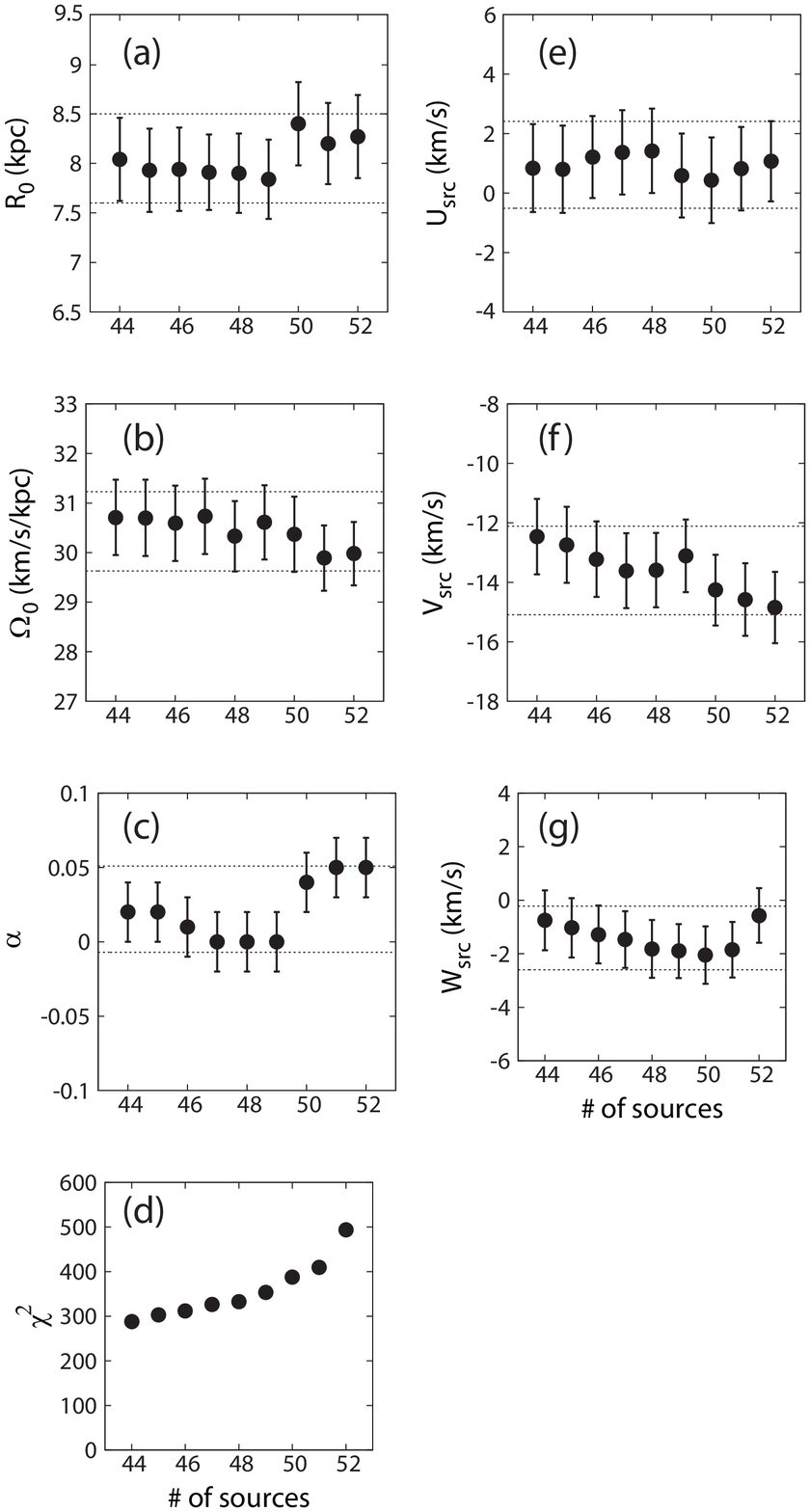}
\end{center}
\caption{Galactic parameters obtained in results 1--9 in table 3. From
 top left to bottom (a to d): for $R_0$, $\Omega_0$, $\alpha$ and $\chi^2$. 
 From top right to bottom (e to g): for $U_{\rm src}$, $V_{\rm src}$,
 and  $W_{\rm src}$. Error
 bars of each point show statistical uncertainties obtained with the MCMC
 analyses (1$\sigma$). All figures are plotted as a function of the number of sample
 sources.
 Dotted lines show the range of our best estimates including systematics
 introduced by different samples with/without possible {``}outliers{''}.}
\end{figure*}

\begin{table*}
\begin{center}
\caption{MCMC results of the Galactic parameters for various sample sets}
\begin{tabular}{cccccccccc}
\hline\hline
 result & \# of & $R_0$ & $\Omega_0$  & $\alpha$ & $U_{\rm src}$ & $V_{\rm src}$ &
 $W_{\rm src}$ & $V_\odot$ & $\chi^2/{\rm d.o.f.}$\\ 
  ID & source  & (kpc) & (km/s/kpc) &  &
 (km~s$^{-1}$) & (km~s$^{-1}$) & (km~s$^{-1}$) & (km~s$^{-1}$) & \\ 
\hline
1 & 52 & 8.27$\pm$0.42 & 29.98$\pm$0.64 & $+$0.05$\pm$0.02 & 1.1$\pm$1.3 &
 $-$14.8$\pm$1.2 & $-$0.6$\pm$1.0 & 5.25$^*$ & 494/150 \\
2 & 51 & 8.20$\pm$0.41 & 29.89$\pm$0.66 & $+$0.05$\pm$0.02 & 0.8$\pm$1.4 &
 $-$14.6$\pm$1.2 & $-$1.9$\pm$1.0 & 5.25$^*$ & 410/147 \\
3 & 50 & 8.40$\pm$0.42 & 30.37$\pm$0.76 & $+$0.04$\pm$0.02 & 0.4$\pm$1.4 &
 $-$14.3$\pm$1.2 & $-$2.0$\pm$1.1 & 5.25$^*$ & 388/144 \\
4 & 49 & 7.84$\pm$0.40 & 30.61$\pm$0.75 & $+$0.00$\pm$0.02 & 0.6$\pm$1.4 &
 $-$13.1$\pm$1.2 & $-$1.9$\pm$1.0 & 5.25$^*$ & 353/141 \\
5 & 48 & 7.90$\pm$0.40 & 30.33$\pm$0.71 & $+$0.00$\pm$0.02 & 1.4$\pm$1.4 &
 $-$13.6$\pm$1.2 & $-$1.8$\pm$1.1 & 5.25$^*$ & 333/138 \\
6 & 47 & 7.91$\pm$0.38 & 30.73$\pm$0.76 & $+$0.00$\pm$0.02 & 1.4$\pm$1.4 &
 $-$13.6$\pm$1.3 & $-$1.5$\pm$1.1 & 5.25$^*$ & 327/135 \\
7 & 46 & 7.94$\pm$0.42 & 30.59$\pm$0.76 & $+$0.01$\pm$0.02 & 1.2$\pm$1.4 &
 $-$13.2$\pm$1.3 & $-$1.3$\pm$1.1 & 5.25$^*$ & 312/132 \\
8 & 45 & 7.93$\pm$0.42 & 30.70$\pm$0.77 & $+$0.02$\pm$0.02 & 0.8$\pm$1.5 &
 $-$12.7$\pm$1.3 & $-$1.0$\pm$1.1 & 5.25$^*$ & 303/129 \\
9 & 44 & 8.04$\pm$0.42 & 30.71$\pm$0.76 & $+$0.02$\pm$0.02 & 0.8$\pm$1.5 &
 $-$12.5$\pm$1.3 & $-$0.8$\pm$1.1 & 5.25$^*$ & 288/126 \\
\hline
\end{tabular}
\end{center}
* : fixed parameter\\

For result 1, all 52 sources are used. For other samples, following
 sources are removed one-by-one from result 2 to 9:
 IRAS 20126+4104, G48.61$+$0.02, G9.62$+$0.20, T-Tau/Sb, AFGL 2591,
 NGC 7538, IRAS 06061+2151, and AFGL 2789.
Error bars show statistical uncertainties obtained with the MCMC analyses.
\end{table*}

\begin{table*}
\begin{center}
\caption{Correlation coefficients between Galactic parameters for result
 4 (49 sources)} 
\begin{tabular}{crrrrrr}
\hline\hline
  & \multicolumn{1}{c}{$R_0$} & \multicolumn{1}{c}{$\Omega_0$} & \multicolumn{1}{c}{$U_{\rm src}$} & \multicolumn{1}{c}{$V_{\rm src}$} &
 \multicolumn{1}{c}{$W_{\rm src}$} & \multicolumn{1}{c}{$\alpha$} \\
\hline
$R_0$         & 1.000 &$-0.164$	&  0.096 &$-0.397$&  0.065 &  0.494 \\
$\Omega_0$    &       & 1.000	&$-0.576$&$-0.052$&$-0.006$&$-0.048$ \\
$U_{\rm src}$ &       &       	&  1.000 &  0.027 &  0.023 &  0.002 \\
$V_{\rm src}$ &       &       	&        &  1.000 &$-0.031$&$-0.148$ \\
$W_{\rm src}$ &       &       	&        &        &  1.000 &  0.022 \\
$\alpha     $ &       &       	&        &        &        &  1.000 \\
\hline
\end{tabular}
\end{center}
\end{table*}

\begin{table*}
\begin{center}
\caption{MCMC results of the Galactic parameters for different model parameters}
\begin{tabular}{cccccccccc}
\hline\hline
 result & \# of & $R_0$ & $\Omega_0$  & $\alpha$ or ($a_0$, $b_0$)$^c$ & $U_{\rm src}$ & $V_{\rm src}$ &
 $W_{\rm src}$ & $V_\odot$ & $\chi^2/{\rm d.o.f.}$\\ 
  ID & source & (kpc) & (km/s/kpc) &  &
 (km~s$^{-1}$) & (km~s$^{-1}$) & (km~s$^{-1}$) & (km~s$^{-1}$) & \\ 
\hline
10 & 52 & 8.27$\pm$0.42 & 29.98$\pm$0.64 & $+$0.05$\pm$0.02 & 1.1$\pm$1.3 &
 $-$14.8$\pm$1.2 & $-$0.6$\pm$1.0 & 5.25$^*$ & 494/150 \\
11 & 52 & 8.25$\pm$0.40 & 29.15$\pm$0.63 & $+$0.05$\pm$0.02 & 1.1$\pm$1.4 &
 $-$8.1$\pm$1.2 & $-$0.5$\pm$1.0 & 12.0$^*$ & 497/150 \\
12 & 52 & 8.25$\pm$0.40 & 28.10$\pm$0.66 & $+$0.05$\pm$0.02 & 1.3$\pm$1.4 &
 0.0$^*$ & $-$0.5$\pm$1.0 & 19.9$\pm$1.2 & 499/150\\
\hline
13 & 49 & 7.84$\pm$0.40 & 30.61$\pm$0.75 & $+$0.00$\pm$0.02 & 0.6$\pm$1.4 &
 $-$13.1$\pm$1.2 & $-$1.9$\pm$1.0 & 5.25$^*$ & 353/141 \\
14 & 49 & 7.82$\pm$0.41 & 29.60$\pm$0.74 & $+$0.00$\pm$0.02 & 0.8$\pm$1.4 &
 $-$6.3$\pm$1.2 & $-$1.9$\pm$1.1 & 12.0$^*$ & 353/141 \\
15 & 49 & 7.77$\pm$0.38 & 28.89$\pm$0.71 & $+$0.00$\pm$0.02 & 0.7$\pm$1.4 &
 0.0$^*$ & $-$1.9$\pm$1.1 & 18.2$\pm$1.2 &  357/141 \\
\hline
16 & 46 & 7.94$\pm$0.42 & 30.59$\pm$0.76 & $+$0.01$\pm$0.02 & 1.2$\pm$1.4 &
 $-$13.2$\pm$1.3 & $-$1.3$\pm$1.1 & 5.25$^*$ & 312/132 \\
17 & 46 & 7.93$\pm$0.40 & 29.74$\pm$0.74 & $+$0.01$\pm$0.02 & 1.3$\pm$1.4 &
 $-$6.5$\pm$1.3 & $-$1.3$\pm$1.1 & 12.0$^*$ & 314/132 \\
18 & 46 & 7.90$\pm$0.38 & 28.86$\pm$0.76 & $+$0.01$\pm$0.02 & 1.3$\pm$1.4 &
 0.0$^*$ & $-$1.3$\pm$1.1 & 18.3$\pm$1.2 & 315/132 \\
\hline
19 & 49$^{a}$ & 7.68$\pm$0.28 & 30.61$\pm$0.53 & $-$0.00$\pm$0.02 & 0.5$\pm$1.0 &
 $-$12.8$\pm$0.9 & $-$1.8$\pm$0.8 & 5.25$^*$ & 603/141 \\
20 & 49$^{b}$ & 7.97$\pm$0.51 & 30.50$\pm$0.97 & $+$0.01$\pm$0.03 & 1.0$\pm$2.0 &
 $-$13.3$\pm$1.7 & $-$1.9$\pm$1.5 & 5.25$^*$ & 192/141 \\
\hline
21 & 49 & 7.72$\pm$0.41 & 30.58$\pm$0.71 & $-$0.1$\pm$0.7 & 0.6$\pm$1.4 &
 $-$13.2$\pm$1.3 & $-$1.9$\pm$1.0 & 5.25$^*$ & 361/140 \\
   &     &             &              & 0.1$\pm$0.2 &    &
  & & & \\
22 & 49 & 7.70$\pm$0.40 & 29.71$\pm$0.71 & $-$0.1$\pm$0.7 & 0.7$\pm$1.4 &
 $-$6.5$\pm$1.3 & $-$1.9$\pm$1.1 & 12.0$^*$ & 367/140 \\
   &     &             &              & 0.1$\pm$0.2 &    &
  & & & \\
23 & 49 & 7.70$\pm$0.40 & 28.91$\pm$0.71 & $-$0.1$\pm$0.7 & 0.6$\pm$1.4 &
  0.0$^*$ & $-$1.9$\pm$1.1 & 18.3$\pm$1.3 & 364/140 \\
   &     &             &              & 0.1$\pm$0.2 &    &
  & & & \\
\hline
\end{tabular}
\end{center}
* : fixed parameter\\
a : with $\sigma_v=5$ km~s$^{-1}$ instead of 7 km~s$^{-1}$. \\
b : with $\sigma_v=10$ km~s$^{-1}$ instead of 7 km~s$^{-1}$. \\
c : through models 10 to 20, $\alpha$ (non-dimensional RC index), and for
 models 21--23, upper raw corresponds to $a_0=d\Theta/dR$ (km/s/kpc),
 and lower raw corresponds to $b_0=(d^2\Theta/dR^2)/2$
 (km/s/kpc$^2$), respectively (values defined at $R=R_0$)
\end{table*}

In order to obtain the best estimates of the Galactic parameters, we
 conducted several MCMC analyses for different sample data sets as well as
 different set of model parameters.
First in section 5.1, we present the results with different samples
 (while fixing model parameters), and then in section 5.2 we treat cases
 with different model parameters.

\subsection{Results with varying number of sample sources}

Table 3 summarizes the MCMC results for different samples, with the
 source number varying from 52 to 44.
Here in table 3, the model parameters are Galactic constants $R_0$ and
 $\Omega_0$, the power-law index of the rotation curve, $\alpha$, the
 mean SFRs motions (with respect to the Galactic rotation)
 $U_{\rm src}$, $V_{\rm src}$ and $W_{\rm src}$.
Note that the Solar motion $V_\odot$ is fixed to 5.25 km $^{-1}$, adopting the
 results by Dehnen \& Binney (1998).
The effect of varying $V_\odot$ and the rotation curve shape will be
 examined later.
Error bars in table 3 show statistical error (1-$\sigma$) obtained with
 the MCMC analyses.
In table 4, we also show the correlation coefficients between Galactic
 parameters for the sample of 49 sources as an example (result 4 in
 table 3).
As seen in table 4, some correlation coefficients exceed 0.2,
 indicating relatively large correlation between parameters.
Part of such correlation can be seen, for instance, in figures 2a, 2c
 and 2f, where cases with higher $R_0$ for results 1--3 provide
 higher $\alpha$ and lower $V_{\rm src}$,
but note that the results presented here are determined including the
 effect of such correlation.
Of course, the strongest correlation exists between $V_\odot$ and $V_{\rm
 src}$ (which is not shown in table 4), which prevents us from determining
 these two parameters at the same time.

As is described in section 2, the full source list consists of 52 sources
for which results of precise astrometry are available.
However, in the sample sources, there likely exist {``}outliers,{''} which have
large deviations from the model-predicted radial velocity and/or proper
motions.
To investigate the effect of such outliers, we ran MCMC analyses
for several sets of sample sources by eliminating possible outliers one-by-one.
The elimination was done based on the conditional probability 
$P_i(\mathbf{O}_i|\mathbf{M})$: a new sample
 of $N-1$ sources was created by eliminating the source with the smallest
 value of $P_i(\mathbf{O}_i|\mathbf{M})$ in the MCMC results for the
 $N$-source sample.
In table 3 (and 5) we also provide $\chi^2$ and degrees of
freedom (d.o.f.) for each model (rather than
 the likelihood), because they are fairly useful to see the effect of
 outliers.

As we will see later in this section (e.g, table 5), the parameters
 related to the Galactic rotation velocity such as $\Omega_0$, 
$V_{\rm src}$ and $V_\odot$ are tightly correlated with each other, and thus
 will be treated in further detail in next subsection (by solving for
 $V_{\rm src}$ or $V_\odot$).
Other parameters such as $R_0$, $\alpha$, $U_{\rm
 src}$, and $W_{\rm src}$ are independent of the choice of model
 parameters, and thus the results listed in table 3 give strong
 constraints on these parameters.
In figures 2, we show resultant model parameters with the number of
 sample sources $N_{\rm src}$ varying from 52 to 44: $R_0$, $\Omega_0$,
 $\alpha$, $U_{\rm src}$, $V_{\rm src}$, $W_{\rm src}$, and $\chi^2$
 versus the number of the sources $N_{\rm src}$.
For reference, the sources with small conditional probability
 are also labeled in table 1.
The effects of {``}outliers{''} can been seen in figures 2. 
For instance, in figure 2d the value $\chi^2$ rapidly increases beyond 50, and
 $\chi^2$ of the full sample (52 objects) is nearly 500 while it is
around 350 for 49 objects (see also results 1 and 4 in table 3).
Therefore, it is likely that at least these three sources (IRAS
 20126+4104, G48.61+0.02, G9.62+0.20) are indeed outliers, and these
 three sources may be better to be excluded from the sample to obtain
 the best estimates of the Galactic parameters.
In fact, previous studies reported large peculiar motions for these
 sources: 
G48.61+0.02 was found to be associated with supernova W51C (Nagayama et
 al. 2011), and thus the peculiar motion is likely to be originated from a
 supernova explosion.
Sanna et al.(2009) suggested that large deviation ($\sim 40$
 km~s$^{-1}$) of G9.62+0.02 from the Galactic rotation may be due to the
 bar potential in the inner Galaxy.
The cause of relatively large non-circular motion of IRAS 20126+4104 is
 unknown, but Moscadelli et al.(2011) already pointed out an existence
 of that at $\sim$20 km~s$^{-1}$ level when a standard rotation curve is
 considered.

Due to the effect of these outliers, the best-estimate values of
some parameters such as $R_0$, $\alpha$, etc. indeed vary with decreasing
$N_{\rm src}$ from 52 to 49 (as in figures 2).
On the other hand, reducing $N_{\rm src}$ further (below 48)
do not alter the best-estimates of those parameters as much.
Therefore, one can expect that the sample with $\sim$48 sources
do not contain {``}outliers.{''}


Although at least the top-three sources in terms of low conditional
probability are highly likely to be outliers,
here we do not simply eliminate these outliers from our samples, but 
we take a rather conservative approach by taking both samples
with/without outliers into account and consider the effects of outliers
as possible systematic error in our analyses.
For instance, in figure 2a, the Galaxy center distance $R_0$ is
determined with an statistical uncertainty $\sigma_{\rm
stat}$ of 0.4 kpc (e.g., for $N_{\rm src}\sim 48$).
On the other hand, the best estimated of $R_0$ for $N_{\rm src}=50$ and
52 differs from that for $N_{\rm src}=48$ slightly beyond the
statistical error bar.
This clearly demonstrates the systematic effect of outliers, and we
regard this scatter as the systematic error in our estimates of the 
Galactic parameters.
In practice, we obtain the best estimate of the parameters by
taking the mean of the results with the sample source number varying
from 44 to 52, and also calculate its standard deviation as the
systematic error $\sigma_{\rm sys}$.
Then we evaluate the final uncertainty by summing up in quadrature both
typical statistical error (taken as the mean of 
statistical error bars for $N_{\rm src}$ ranging from 44 to
52) and systematic error $\sigma_{\rm sys}$.
Note that here the selection of the smallest number of the
sample sources to be used is rather arbitrary, but in the present
analyses we adopt 44 for the minimum number of the sources.
In fact lowering the number of the sample further would not alter the
 resultant parameters drastically, but the uncertainty would be slightly
 increasing due to statistical effect.

The best estimates of Galactic parameters obtained in that way are also
plotted in each panel of figures 2 by dotted lines showing the possible
range.
For the best estimate of $R_0$, we obtained
\begin{equation}
R_0=8.05\pm0.45 \;\;{\rm kpc}.
\end{equation}
This is 5\%-level determination of the most fundamental Galactic
 constant $R_0$.
As we will see in discussion later, this is consistent with other recent
 studies of $R_0$ such as using motions of stars around the Galaxy center
 or variable stars, but we note that our result is independent
 of others and provides another constraint on the distance to the Galaxy
 center.

The rotation curve index $\alpha$ is also tightly constrained to be
\begin{equation}
\alpha=0.022\pm0.029. 
\end{equation}
The index is 0 within the error bar, confirming
that the rotation curve of the Galaxy is basically flat in the regions
where our samples are distributed (Galacto-centric radius of 4 to 13 kpc).
The use of the full sample gives a marginally positive value of $\alpha$
of 0.05, but we note that $\alpha=0.05$ results in a rise of rotation
velocity only by $\pm$3 \% between Galacto-centric radius between 5 and
15 kpc.
Hence, even in the maximum case of $\alpha$, Galactic rotation curve
flatness is $\pm$3\% level in those regions.

The $U$ and $W$ components of the source non-circular motions are also
well constrained here, and we obtained 
\begin{equation}
 U_{\rm src}=0.95\pm 1.46 \;\; {\rm km~s}^{-1},
\end{equation}
\begin{equation}
 W_{\rm src}=-1.41\pm 1.19 \;\;{\rm km~s}^{-1}.
\end{equation}

The mean inward motion $U_{\rm src}$ is obtained to be 0 within the
error bar.
Also, the mean perpendicular motion $W_{\rm src}$ is found to be
practically 0 within the error bars, being independent of the models.
This is rather natural when one considers the symmetry of Galactic disk
along the Galactic latitude.

These parameters do not depend on the various models considered in the next
section, and we take them as our final results for these parameters.
Regarding other parameters, from figures 2 we also obtained
 $\Omega_0=30.43\pm0.80$ km~s$^{-1}$ kpc$^{-1}$, $V_{\rm src}=-13.60\pm
 1.49$ km~s$^{-1}$, but as will be seen in the next subsection, we have
 to consider other systematic effects to obtain the final results.

\subsection{Results with changing model parameters}

In this subsection we present the MCMC results of the Galactic
parameters by choosing different model parameter sets and also different
rotation curves.
Table 5 summarizes the results, again presenting results of the best
estimates of the parameters, as those in table 3.
In table 5 results 10--12 are a family of the best estimates for the full
sample of 52 sources, with
different treatment of $V_{\rm src}$ and $V_\odot$: in result 10 we
adopt $V_\odot$=5.25 km $^{-1}$ (Dehnen \& Binney 1998, note
that result 10 is identical to result 1 in table 3), while in result 11
$V_\odot$=12.0 km $^{-1}$ (Sch\"{o}nrich et al. 2010) is adopted. 
In both results 10 and 11, $V_{\rm src}$ is considered as a parameter to
be solved for.
As seen in results 10 and 11, the SFRs lag velocities (negative $V_{\rm
src}$) differ by nearly the same amount with the change in adopted
$V_\odot$, indicating strong correlation between $V_{\rm src}$ and $V_\odot$.
On the other hand, in result 12, instead we fixed $V_{\rm src}$=0 km
s$^{-1}$ (no SFRs lag) and solved for the Solar motion $V_\odot$.
In other words, in result 12, the apparent slow rotation of SFRs is totally
attributed to the Solar peculiar motion.
Results 13--15, and 16--18 are other families of the results similar to
results 10--12,  but for different numbers of the sample sources:
results 13--15 are for the sample of 49 sources, and results
16--18 are for the sample of 46 sources (note that results 4 and 13, and
results 7 and 16 are identical to each other, respectively).
The elimination of sources from the full sample is conducted based on
the conditional probability $P_i(\mathbf{O}_i|\mathbf{M})$
, as is already described in 5.1.
These results with different number of sources are again presented to evaluate
the effect of samples with/without outliers.

Results 19 and 20 are to investigate the dependence of the results on
the adopted value of 1-dimensional velocity dispersion of SFRs
($\sigma_{v}$): throughout the paper, $\sigma_{v}$=7 km~s$^{-1}$ is
adopted, but only for results 19 and 20, this is altered to 5
or 10 km~s$^{-1}$, respectively ($V_\odot$=5.25 km
s$^{-1}$ is adopted for both results 19 and 20).
Results 19 and 20 should be compared with results 13, because they are based
on the same sample and parameters with only difference in $\sigma_v$.
As can be seen in table 5, the results 19, 20 and 13 are fairly
consistent with each other.
The only difference is the size of statistical error bars and
the value of $\chi^2$,  as is expected in case of using different
$\sigma_v$ : result 19 gives $\chi^2$
of 603 against d.o.f. of 141, and result 20 gives $\chi^2$ of 192.
They are to be compared with result 13 provides $\chi^2$ of 353.
In terms of reduced $\chi^2$ ($=\chi^2/{\rm d.o.f}$), $\sigma_v$ of 10
km~s$^{-1}$ seems too large, and our choice of $\sigma_v=7$ km~s$^{-1}$
seems more reasonable.
The reduced $\chi^2$ may indicate that $\sigma_v=5$ km~s$^{-1}$ is
reasonbale, but here we conservatively adopt $\sigma_v=7$ km~s$^{-1}$
because the use of too small $\sigma_v$ could lead to underestimation of
statistical error.
Note that the adopted value of $\sigma_v=7$ km~s$^{-1}$ is similar to
those in the previous studies: for instance, Reid et al.(2009b) used 7 km
s$^{-1}$ for $\sigma_v$, and McMillan \& Binney (2010) solved $\sigma_v$
by setting it as a parameter to be estimated, and obtained $\sigma_v$
from 7 to 10 km~s$^{-1}$ depending on their models.

The last three results (21--23) in table 5 are those for using different
rotation curve models.
Reid et al.(2009b) and McMillan \& Binney (2010) suggested that the Galactic
parameters also depend on choice of rotation curve shape.
In most cases in table 3 and 5 we used a power-law rotation curve.
In order to see how results would be altered with a different rotation
curve, we adopt the 2nd-order-polynomial rotation curve
in equation (\ref{eq:2pol-RC}) for results 21-23.
This rotation curve 
has two parameters, while a power-law rotation curve has only one parameter.
However, as can be seen in table 5, the use of a different rotation curve
does not change our results drastically.
This may indicate that the shape rotation curve determined with the
current sample is basically
well-represented by a simple function such as the one-parameter power-low model.

From table 5, one can see that the parameters $R_0$, $\alpha$,
$U_{\rm src}$ and $W_{\rm src}$ are not strongly dependent on the
choice of $V_\odot$ or $V_{\rm src}$.
For instance, from results 11 to 20, $R_0$ ranges from 7.7 to 8.3
kpc, but basically this is the sample effect: in fact, the best
estimate of $R_0$ remains unchanged within a family of results based on
the same sample (such as results 10--12, 13--15 etc.).
The variation of $R_0$ for results 10--23 thus originates from the
difference in the number of sources (and hence inclusion/exclusion of
outliers).
The best estimate of $R_0$ obtained in 5.1, which is $R_0$=8.05$\pm$0.45 kpc
including systematic error, fully covers the range of $R_0$
in table 5.
Therefore, as we described in the previous subsection, we can consider the
value of $R_0$ obtained in the previous subsection as the final value for the
current study.
In addition to $R_0$, the same applies to $\alpha$, $U_{\rm src}$ and
$W_{\rm src}$, and the best estimates obtained in section 5.1 cover the full
 ranges of these parameters in table 5.

\begin{figure}[t]
\begin{center}
\hspace*{-0.5cm}
       \FigureFile(85mm,85mm){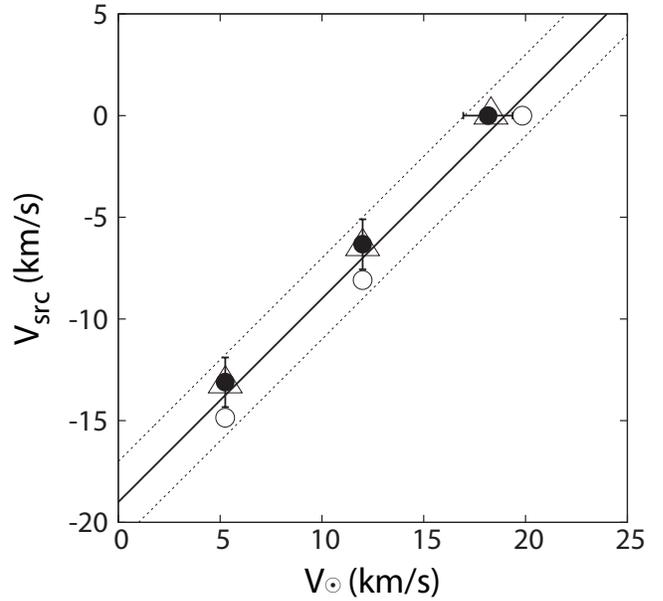}
\end{center}
\caption{Correlation between $V_\odot$ and $V_{\rm src}$. Open circles
 are those for results 10--12 in table 5, filled circles with error bars
 for results 13--15, and open triangles for results 16--18,
 respectively.
Thin line corresponds to the best fit relation of $V_{\rm src}=V_\odot -
 19 (\pm 2)$ km~s$^{-1}$, with dotted lines showing the error bar range.}
\end{figure}
\begin{figure}[h]
\begin{center}
\hspace*{-0.5cm}
       \FigureFile(85mm,85mm){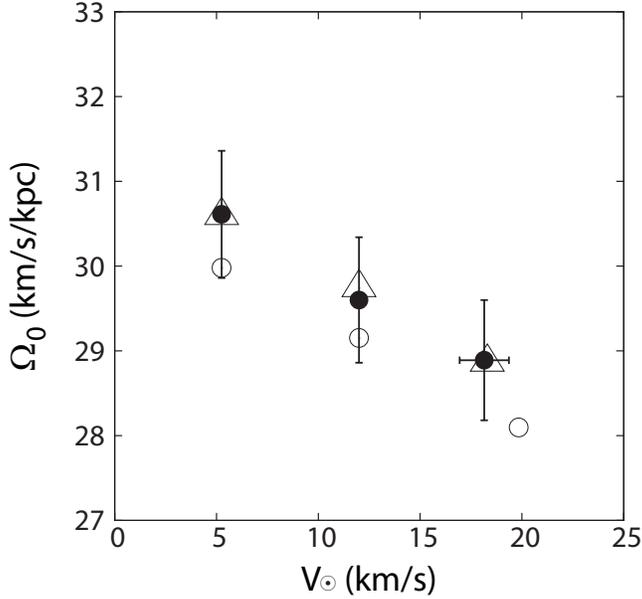}
\end{center}
\caption{Correlation between $V_\odot$ and $\Omega_0$. Open circles
 are those for results 10--12 in table 5, filled circles with error bars
 for results 13--15, and open triangles for results 16--18,
 respectively.}
\end{figure}

On the other hand, in table 5, 
there are strong correlations between the Solar peculiar
motion $V_\odot$, the mean source motion in the rotational direction 
$V_{\rm src}$, and the angular velocity of the LSR $\Omega_0$.
One can see this correlation by comparing the results for the same
sample but with different $V_\odot$ or $V_{\rm src}$ (e.g., results 10--12).
To clearly show this situation, in figure 3 we show the relation between
$V_\odot$ and $V_{\rm src}$ for the results in table 5.
The open circles in figure 3 are for results 10--12, filled
circles (with error bars) for results 13--15, and open triangles for
results 16--18, respectively.
As seen in figure 3, the relation between $V_\odot$ and
$V_{\rm src}$ can be approximated by a linear function, with 
$V_{\rm src}=V_\odot - 19$ km~s$^{-1}$ (thin line in figure 3).
When one compares $\chi^2$ values with the three results based on
the same sample (e.g., models 10--12, 13--15, ...), there is
little difference in $\chi^2$, indicating that our results do not show
any preference for a particular choice of $V_{\rm src}$ and $V_\odot$.
Therefore, at this stage, the only thing we can constrain is that these
two quantities must satisfy the following relation
\begin{equation}
\label{eq:Vsrc-Vsun}
 V_{\rm src} = V_\odot - 19\, (\pm 2) \,\,{\rm km~s}^{-1}
\end{equation}
The allowed region of ($V_\odot$, $V_{\rm src}$) with the above error
bar ($\pm 2$ km~s$^{-1}$) is also shown in figure 3 with dotted lines.
If $V_\odot=5.25$ km s$^{-1}$ is adopted to this relation,
 the star-forming regions lag behind the Galactic rotation by $\sim14$
 km~s$^{-1}$, confirming the earlier suggestion by Reid et al.(2009b).
On the other hand, an upward modification of $V_\odot$
 by Sch\"{o}nrich et al. (2010) with $V_\odot=12.0$ km~s$^{-1}$ reduces
 the lag down to $\sim 7$ km s$^{-1}$ level.
Only $V_\odot$ of $\sim 20$ km s$^{-1}$ can lead to no lag case, but
this solar motion is larger than the classical IAU value of the Solar
motion, with $|V|=20$ km~s$^{-1}$ toward (18$^h$, 30$^\circ$) in B1900
(e.g., Kerr \& Lynden-Bell 1986), corresponding to $V_\odot=15$
km~s$^{-1}$.
Hence, this should be taken to be an extreme case, and probably reality
would lie among these cases (e.g., up-ward modified $V_\odot$ with some
rotation lag).
The tight correlation between $V_\odot$ and $V_{\rm src}$ can be simply
attributed to the fact that the observed sources motions are relative to
that of the Solar system and hence the effect of the Solar motion and
lag of star-forming regions are still degenerated.

\begin{figure}[t]
\begin{center}
\hspace*{-0.8cm}
       \FigureFile(85mm,85mm){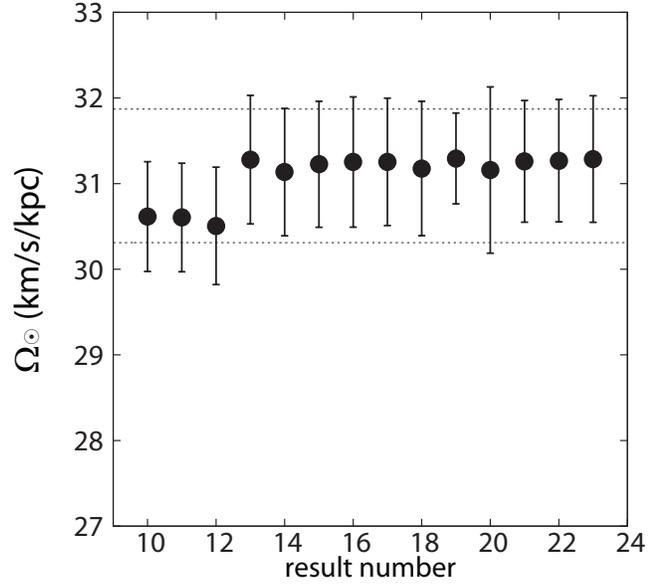}
\end{center}
\caption{Angular rotation velocity of the Sun, $\Omega_\odot$, obtained
 for results 10--23 in table 5 (filled circles with error bars).
Horizontal axis is the result ID, running from 10 to 23.
Dotted lines show the possible range of $\Omega_\odot$ for our best
 estimate, $\Omega_\odot = 31.09 \pm 0.78$ km~s$^{-1}$ kpc$^{-1}$.
}
\end{figure}

\begin{table}[t]
\begin{center}
\caption{Additional Galactic parameters}
\begin{tabular}{ccccc}
\hline\hline
model & $V_\odot$ & $V_\odot - V_{\rm src}$ & $\Omega_\odot$ & $\Theta_0$ \\
  ID  & km~s$^{-1}$ & km~s$^{-1}$ & km~s$^{-1}$ kpc$^{-1}$ & km~s$^{-1}$ \\
\hline
10  &  5.25    & 20.1$\pm$1.2          & 30.61$\pm$0.64 & 247.9$\pm$13.7 \\
11  &  12.0    & 20.1$\pm$1.2          & 30.60$\pm$0.63 & 240.5$\pm$12.8 \\
12  &   *      & 19.9$\pm$1.2          & 30.50$\pm$0.69 & 231.8$\pm$12.5 \\
\hline
13  &  5.25    & 18.4$\pm$1.2          & 31.28$\pm$0.75 & 240.0$\pm$13.6 \\
14  &  12.0    & 18.3$\pm$1.2          & 31.13$\pm$0.74 & 231.5$\pm$13.4 \\
15  &   *      & 18.2$\pm$1.2          & 31.23$\pm$0.74 & 224.5$\pm$12.3 \\
\hline
16  &  5.25    & 18.5$\pm$1.3          & 31.25$\pm$0.76 & 242.9$\pm$14.2 \\
17  &  12.0    & 18.5$\pm$1.3          & 31.25$\pm$0.74 & 235.8$\pm$13.3 \\
18  &   *      & 18.3$\pm$1.2          & 31.18$\pm$0.78 & 228.0$\pm$12.5 \\
\hline
19  &  5.25    & 18.0$\pm$0.9          & 31.29$\pm$0.53 & 235.1$\pm$ 9.5 \\
20  &  5.25    & 18.5$\pm$1.7          & 31.16$\pm$0.97 & 243.1$\pm$17.4 \\
\hline
21  &  5.25    & 18.5$\pm$1.3          & 31.26$\pm$0.71 & 236.1$\pm$13.7 \\
22  &  12.0    & 18.5$\pm$1.3          & 31.27$\pm$0.71 & 228.8$\pm$13.1 \\
23  &   *      & 18.3$\pm$1.3          & 31.29$\pm$0.74 & 222.6$\pm$12.8 \\
\hline
\end{tabular}
\end{center}
* denotes that $V_\odot$ is solved for by setting $V_{\rm src}=0$ km
 s$^{-1}$. 
The resultant $V_\odot$ is identical to the value in the right column of 
$V_\odot - V_{\rm src}$.
$\Theta_0$ is calculated using the relevant values of $R_0$ and
 $\Omega_0$ in table 5.
\end{table}

The correlation between $V_{\rm src}$ and $V_\odot$ also affects the angular
rotation velocity of the LSR $\Omega_0$.
For instance, $\Omega_0$ simply decreases in results 10--12 as $V_\odot$
increases, and the same trends are found in results 13--15 and 16--18 in
table 5.
Figure 4 illustrates this correlation between $V_\odot$ and $\Omega_0$,
with the same symbols as in figure 3.
Here again $V_\odot$ and $\Omega_0$ show a linear trend.
Considering this situation, it is useful to introduce a following new
Galactic parameter
\begin{equation}
 \Omega_\odot \equiv \Omega_0 + \frac{V_\odot}{R_0},
\end{equation}
 which is the angular velocity of the Sun.
Table 6 summarizes the derived $\Omega_\odot$ and other parameters
such as $V_\odot-V_{\rm src}$ and $\Theta_0$, based on the results in
table 5.
Also, figure 5 shows $\Omega_\odot$ for results 10--23 in table 5.
As seen in figure 5, the resultant $\Omega_\odot$ shows only weak
dependence on the results (number of the source, choice of $V_\odot$
etc.), and thus is readily constrained well in the present study.
As was done for $R_0$ determination in the previous subsection,
from figure 5, $\Omega_\odot$ is determined as
\begin{equation}
 \Omega_\odot = 31.09 \pm 0.78 \; {\rm km\; s}^{-1}{\rm \; kpc}^{-1}.
\end{equation}
Here again the error bar includes systematics due to the effect of outliers. 
This is consistent with the direct measurement of Sgr A* motion
along the Galactic plane carried out by Reid \& Brunthaler (2004), 
who obtained $\mu_l=6.379\pm0.024$ mas yr$^{-1}$, or the angular
rotation velocity $\Omega_\odot$ of 30.24$\pm$0.11 km~s$^{-1}$
kpc$^{-1}$.
Note that while $\Omega_\odot$ is well constrained, in table 6
$\Theta_0$ still remains relatively uncertain.
In fact, in table 6, the full possible range of $\Theta_0$ is
223 -- 248 km $^{-1}$.
Hence the determination of precise $V_\odot$ and $\Theta_0$
is still an important issue for future observations.
In table 7, we summarize final values and relations for Galactic
parameters in the present study.
The error bars in table 7 include both
statistical and systematic error, as already discussed in this section.

\begin{table}[t]
\caption{Summary of Galactic parameters obtained in the present
 paper. Error bars indicated here include systematic errors.}
\begin{center}
\begin{tabular}{ccl}
\hline\hline
$R_0$ & 8.05$\pm$0.45 & kpc \\
$\Omega_\odot$ & 31.09$\pm$0.78 & km~s$^{-1}$ kpc$^{-1}$\\
$\alpha$ & 0.022$\pm$0.029 & \\
$U_{\rm src}$ & 0.95$\pm$1.46 & km~s$^{-1}$ \\
$V_{\rm src}$ & $V_\odot -19\; (\pm2)$ & km~s$^{-1}$ \\
$W_{\rm src}$ & $-1.41\pm1.19$ & km~s$^{-1}$ \\
\hline
\end{tabular}
\end{center}
\end{table}

%

\section{Galactic parameter comparisons}

\subsection{$R_0$}

In the previous section, we have obtained the Galaxy center distance to
be $R_0=8.05\pm0.45 \;\;{\rm kpc}$, with the error bar including
systematics due to possible {``}outliers{''}.
This result is fully consistent with recent studies of $R_0$.
For instance, the review paper by Reid (1993) summarized the previous
studies of $R_0$ and provided $R_0=8.0\pm 0.5$ kpc.
Observations of stellar orbits around Sgr A* also provided the
distance to the Galaxy center as $R_0=8.4\pm0.4$ kpc (Ghez et
al. 2008) and $R_0=8.33\pm0.35$ kpc (Gillessen et al. 2009).
Earlier analysis of Galactic structure based on maser astrometry (Reid
et al. 2009b) also reported $R_0=8.4\pm0.6$ kpc.
Period-Luminosity relation for Mira also suggests
$R_0=8.24\pm0.08|_{\rm stat}\pm0.42|_{\rm sys}$ kpc (Matsunaga et al. 2009).
All these studies are consistent within error bars and thus the result
in the present paper confirms that $R_0$ is between 8 and 8.5 kpc.

\subsection{$\Omega_\odot$ and $\Omega_0$}

The most direct determination of $\Omega_\odot$ is the apparent proper
motion of Sgr A* along the Galactic plane: assuming that the Sgr A* is
at rest at
the center of the Galaxy, its proper motion directly reflects the sum of
the Galactic rotation at the LSR and the Solar motion with respect to
the LSR.
As already mentioned in the previous section, our result of
$\Omega_\odot=31.09\pm0.78$ km s$^{-1}$ kpc$^{-1}$ is consistent with 
the Sgr A* motion obtained  by Reid \& Brunthaler (2004), which was
$\Omega_\odot$ of 30.24$\pm$0.11 km~s$^{-1}$ kpc$^{-1}$.
Note that our astrometry data does not include the proper motion of Sgr A*
itself, and hence these two studies are totally independent.
This consistency assures that the Sgr A* is indeed at rest at the
Galaxy center, with an upper limit of its in-plane motion
less than $\sim 6$ km~s$^{-1}$.
This limit is still larger than the upper limit for the proper motion in
$b$-direction (perpendicular to the plane), which is found to be $\sim
1$ km~s$^{-1}$.
However, the upper limit of Sgr A*'s in-plane motion is significantly
smaller than $\Theta_0$ itself (which ranges 223--248 km~s$^{-1}$ in the
present paper), 
indicating that Sgr A* is truly at the dynamical center of the Galactic
rotation.

Miyamoto \& Zhu (1998) analyzed HIPPARCOS data of OB stars and Cepheid
variables, and derived $\Omega_\odot=32.7\pm1.4$ km~s$^{-1}$ kpc$^{-1}$
from the OB stars and $\Omega_\odot=30.4\pm1.4$ km~s$^{-1}$ kpc$^{-1}$ from
the Cepheids, respectively.
Note that here the results by Miyamoto \& Zhu (1998) are converted to
$\Omega_\odot$ by using their own determination of the Solar
motion, which was $V_\odot \sim 15$ km~s$^{-1}$.
Our result of $\Omega_\odot$ lies in the middle of these two results and
consistent with both of them.
Earlier studies of Galactic parameters based on VLBI astrometry also
resulted in similar values.
For instance, Reid et al.(2009b) obtained $\Omega_0=$30.3$\pm$0.9 km
s$^{-1}$ kpc$^{-1}$ based on $V_\odot=5.25$ km~s$^{-1}$, yielding
$\Omega_\odot=30.9 \pm 0.9$ km~s$^{-1}$ kpc$^{-1}$.
Reanalysis of the same astrometry data also resulted in similar range of
$\Omega_\odot$, with 30.9 to 32.5 km~s$^{-1}$ kpc$^{-1}$ (McMillan \&
Binney 2010) and 32.9$\pm$1.22 km~s$^{-1}$ kpc$^{-1}$ (Bobylev \&
Bajkova 2010).
These results also support that $\Omega_\odot$ slightly larger than 30 km
s$^{-1}$ kpc$^{-1}$.
On the other hand, if one adopts IAU recommendations of Galactic
constants, one would obtain $\Omega_\odot$=(220+15)/8.5=27.6 km~s$^{-1}$
kpc$^{-1}$.
Therefore, the present paper as well as recent other studies on
$\Omega_\odot$ strongly suggest that the Galactic angular velocity
should require an upward modification by 10\% level.
Because $\Theta_0$ is tightly linked to $\Omega_0$, the same applies to
the rotation velocity of the LSR $\Theta_0$:
while the resultant values of $\Theta_0$ are dependent of $V_\odot$ as
well as $R_0$, our results of $\Theta_0$ range between 223 km~s$^{-1}$
and 248 km~s$^{-1}$, being higher than the IAU recommendation of
$\Theta_0$=220 km~s$^{-1}$.

\subsection{Source peculiar motion and $V_\odot$}

As we have found in the previous section, the $U$ and $W$ components of
the mean source peculiar motion are independent of $V_\odot$ and hence well
constrained.
As summarized in table 7, the mean perpendicular motion
$W_{\rm src}(=-1.41\pm1.19$ km s$^{-1}$) can be regarded as 0 at
1.2-$\sigma$ level, 
which is naturally expected from the equilibrium of Galactic disk.
The $U$ component is also 0 within the error bar, and we can safely
conclude that the $U$ component is negligible when compared
to the Galactic rotation velocity, which is an order of $~\sim 240$
km~s$^{-1}$.

On the other hand, as we have seen in the previous section, the $V$
component of the source mean motion is tightly correlated with
$V_\odot$, and thus exact value cannot be fixed based solely on the
VLBI astrometry data.
The only constraint we can obtain is the linear relation between $V_\odot$
and $V_{\rm src}$ given in equation (\ref{eq:Vsrc-Vsun}).
If we adopt $V_\odot=5.25$ km~s$^{-1}$ of Dehnen \& Binney (1998), then
$V_{\rm src}$ becomes $-14\pm$2 km~s$^{-1}$.
Therefore, our result confirms the findings by Reid et al.(2009b) that 
the star-forming regions lag behind the Galactic rotation.
It is remarkable that this confirmation is done based on much larger number
of sources compared to that in Reid et al.(2009b).
However, as we have already seen, models without $V_{\rm src}$ also provide
a reasonable fit to the data: $\chi^2$ only slightly vary with changing
$V_\odot$ or solving for it with assuming $V_{\rm src}=0$ km~s$^{-1}$
(see, e.g., results 12, 15, and 18 in table 5).
Therefore, at this stage we cannot conclude whether or not a Galaxy model with
the rotation lag of the star-forming regions is better than a model with larger
$V_\odot$ and no lag.
However, the strong relation in equation [\ref{eq:Vsrc-Vsun}]
will provide a good base of future investigations on this problem: once one of
the parameters ($V_\odot$ or $V_{\rm src}$) is determined based on other
observations etc., one can immediately find the conclusion on the
(non)existence of the SFRs lag.

\subsection{Rotation Curve}

\begin{figure*}[t]
\begin{center}
       \FigureFile(150mm,150mm){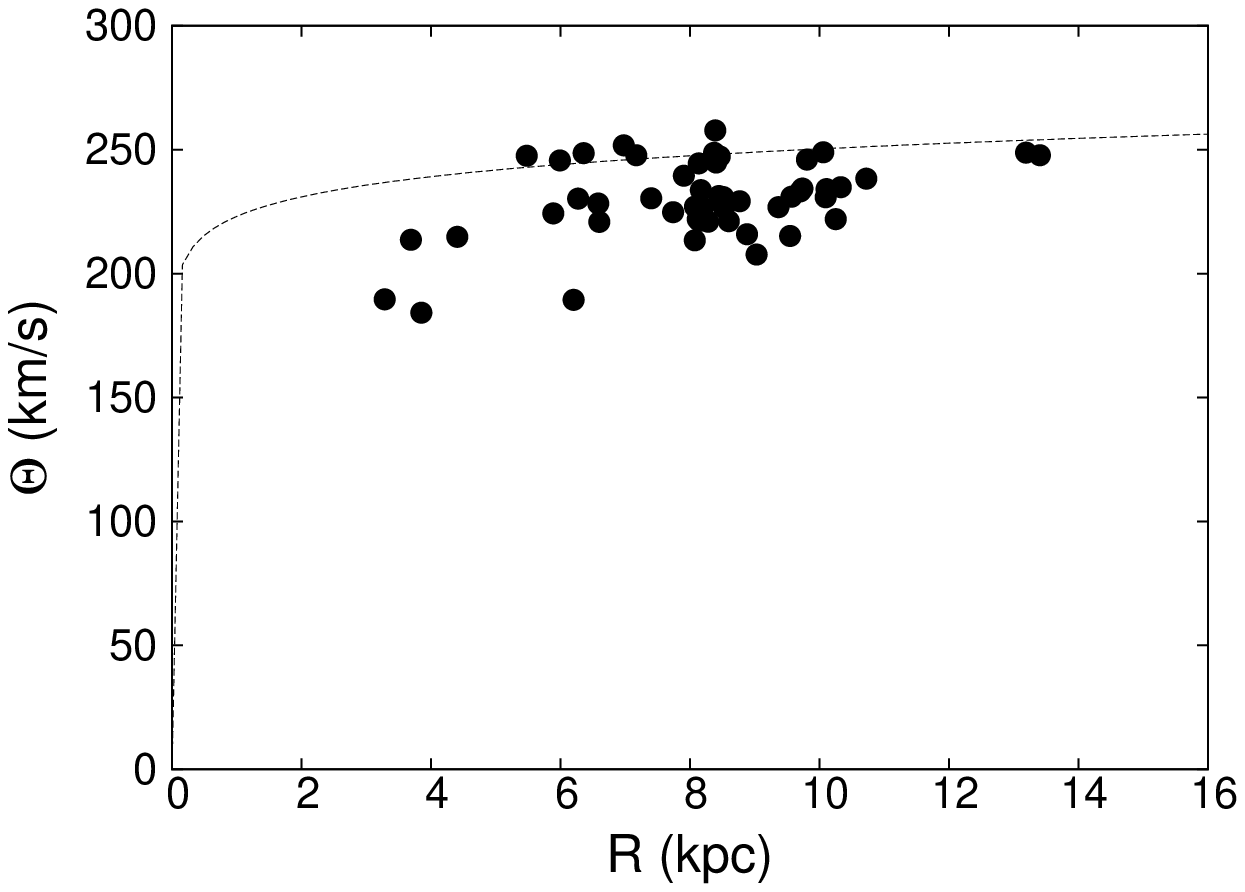}
\end{center}
\caption{Rotation curve plot for result 1 (52 sources). Filled circles
 correspond to the observed sources. Values of $R_0$ and $\Theta_0$ are
 taken from the relevant results in table 1 ($R_0=8.27$ kpc,
 $\Omega_0=29.98$ km s$^{-1}$ kpc$^{-1}$ and thus $\Theta_0=R_0\times \Omega_0 = 248$
 km~s$^{-1}$).
Here error bars are not shown because in most cases they are
 invisibly small in this plot (but note that the systematic effect is
 considerably large as described in the text).
Dotted line shows a rotation curve with $\Theta_0=248$ km~s$^{-1}$ and
 a power low index $\alpha=0.05$.
Note that here $V_\odot=5.25$ km~s$^{-1}$ is adopted and hence the lag of
 SFRs against Galactic rotation is prominent (by $\sim 15$ km~s$^{-1}$),
 with most of data points located below the model rotation curve.}
\end{figure*}

While a few Galactic parameters still have a considerable uncertainty
depending on the Solar motion etc., the shape of the rotation curve
itself is well determined based on the analyses in the previous section.
Figure 6 shows an example of the rotation curve, based on result 1.
Here we adopted from table 3 Galactic parameters of  
$R_0=8.27$ kpc and $\Omega_0 = 29.98$ km~s$^{-1}$ kpc$^{-1}$ (or
$\Theta_0$=248 km~s$^{-1}$), respectively.
As seen in the figure, the rotation curve can be basically regarded as
{``}flat{''} between 4 kpc and 15 kpc, being
consistent with the rotation curves of other galaxies with rotation velocities
comparable to that of the Milky Way Galaxy (e.g., Sofue \& Rubin 2001).

The flat rotation curve is also consistent with previous studies of the
Galactic rotation curves (e.g., Clemens et al. 1985).
On the other hand, Honma \& Sofue (1997) showed that the shape of HI
rotation curve is strongly dependent on $\Theta_0$ and for a flat
rotation curve,
their preferred value of $\Theta_0$ is around 200 km~s$^{-1}$.
This $\Theta_0$ is inconsistent with the results of the VLBI astrometry
here, which suggested $\Theta_0$ much higher than the IAU standard value
of 220 km s$^{-1}$.
For solving this discrepancy, again the Solar motion would be a
clue, because both of the HI rotation curve and Galaxy parameters based
on VLBI astrometry are significantly affected by the Solar motion
$V_\odot$.

The flatness of the rotation curve strongly constrained here
can be used to estimate the distribution of dark matter in
the Galaxy.
In fact, in order to sustain this flat rotation curve, a considerable
amount of dark matter is required.
For instance, suppose that the Galaxy's disk is an exponential disk with
a scale length of $h=3$ kpc and maximum disk rotation 
velocity equal to $\Theta_0$,  and the mass-to-light
ratio is constant through the disk.
Then, as is discussed in Honma et al. (2007), the ratio of disk mass
to the total mass within 13 kpc is around 0.7, and hence 30\% of the
total mass should be in the form of dark matter.

Note that the data points in figure 6 mostly lie below the best-fit
rotation curve.
This is obviously the effect of the lag of the star-forming regions
because in figure 6 we adopt the Solar motion of 5.25 km s$^{-1}$ that
makes the lag largest among the results in table 5.
If larger $V_\odot$ is adopted, then the lag will be
reduced and the data points should scatter around the model curve.
Other interesting feature in figure 6 is that the deviation of the
sources from the rotation curve could be systematic (such as lower
rotation velocity at Galacto-centric radii of 5 kpc and 9 kpc).
Although the number of the sample sources is not enough to discuss this 
at the present, this may be related to non-axisymmetric structure of the
Galaxy (such as a bar and spiral arms), and so this will be interesting
target for future studies with more sources.

\bigskip

We are grateful to anonymous referee for careful reading of the manuscript and
constructive suggestions.
One of the authors (MH) acknowledges the financial support from grant-in-aid
(No.21244019) by the Ministry of Education, Culture, Sports, Science
and Technology (MEXT).
The authors also thank all the staff members at Mizusawa
VLBI observatory for supporting the observations.

\setcounter{table}{0}

\include{table1}

\end{document}

%% file: table1.tex



\begin{table*}

  \caption{Parallaxes and proper motions for 52 sources.}

  \label{tab:1}

  \begin{center}

    \begin{tabular}{lrrrrrrll}

      \hline

      \multicolumn{1}{c}{Source}         &

      \multicolumn{1}{c}{$l$}            &

      \multicolumn{1}{c}{$b$}            &

      \multicolumn{1}{c}{$\pi$}          &

      \multicolumn{1}{c}{$v_{\rm LSR}$}  &

      \multicolumn{1}{c}{$\mu_\alpha \cos \delta$}        &

      \multicolumn{1}{c}{$\mu_\delta$}        &

      \multicolumn{1}{c}{Ref.}           &

      \multicolumn{1}{c}{rank in}          \\

                                           &

      \multicolumn{1}{c}{(deg)}          &

      \multicolumn{1}{c}{(deg)}          &

      \multicolumn{1}{c}{(mas)}          &

      \multicolumn{1}{c}{(km s$^{-1}$)}   &

      \multicolumn{1}{c}{(mas yr$^{-1}$)}  &

      \multicolumn{1}{c}{(mas yr$^{-1}$)}  &

                                          &

      \multicolumn{1}{c}{$P_i(\mathbf{O_i}|\mathbf{M})$}    \\

      \hline

G 5.89$-$0.39		&$	5.89 	$&$	-0.39 	$&$	0.780   \pm	0.050 	$&$	9 	\pm	3 	$&$	0.17	\pm	0.42 	$&$	-0.95 	\pm	0.34 	$&19	& \\

G 9.62+0.20		&$	9.62 	$&$	0.20 	$&$	0.194   \pm	0.023 	$&$	2 	\pm	5 	$&$	-0.58	\pm	0.05 	$&$	-2.49 	\pm	0.27 	$&28	& 3\\

G 12.89+0.49		&$	12.89 	$&$	0.49 	$&$	0.428   \pm	0.022 	$&$	40 	\pm	3 	$&$	0.16	\pm	0.03 	$&$	-1.90 	\pm	1.59 	$&38	& \\

G 14.33$-$0.64		&$	14.33 	$&$	-0.64 	$&$	0.893 	\pm	0.101 	$&$	22 	\pm	10 	$&$	0.95 	\pm	2.00 	$&$	-2.50 	\pm	2.00 	$&31	&\\

G 15.03$-$0.68 		&$	15.03 	$&$	-0.68 	$&$	0.505 	\pm	0.033 	$&$	23 	\pm	3 	$&$	0.68 	\pm	0.05 	$&$	-1.42 	\pm	0.09 	$&38	&\\

G 23.01$-$0.41		&$	23.01 	$&$	-0.41 	$&$	0.218 	\pm	0.017 	$&$	82 	\pm	3 	$&$	-1.72 	\pm	0.04 	$&$	-4.12 	\pm	0.30 	$&4	&\\

G 23.44$-$0.18		&$	23.44 	$&$	-0.18 	$&$	0.170 	\pm	0.032 	$&$	98 	\pm	3 	$&$	-1.93 	\pm	0.10 	$&$	-4.11 	\pm	0.07 	$&4	&\\

G 23.66$-$0.13		&$	23.66 	$&$	-0.13 	$&$	0.313 	\pm     0.039 	$&$	83 	\pm	2 	$&$	-1.32 	\pm     0.02 	$&$	-2.96 	\pm	0.03 	$&3	& \\

G 27.36$-$0.16		&$	27.36 	$&$	-0.16 	$&$	0.125 	\pm     0.042 	$&$	92 	\pm	3 	$&$	-1.81 	\pm     0.08 	$&$	-4.11 	\pm	0.26 	$&38	&\\

EC 95 (Serpens) 	&$      31.56   $&$	5.33 	$&$	2.410 	\pm	0.020 	$&$	 9 	\pm	3 	$&$	 0.70 	\pm	0.02 	$&$	-3.64 	\pm	0.10 	$&6	&\\
G 35.20$-$0.74		&$	35.20 	$&$	-0.74 	$&$	0.456 	\pm	0.045 	$&$	28 	\pm	3 	$&$	-0.18 	\pm	0.06 	$&$	-3.63 	\pm	0.11 	$&39	&\\

G 35.20$-$1.74		&$	35.20 	$&$	-1.74 	$&$	0.306 	\pm	0.045 	$&$	42 	\pm	3 	$&$	-0.71 	\pm	0.05 	$&$	-3.61 	\pm	0.17 	$&39	&\\

G 48.61+0.02		&$	48.61 	$&$	0.02 	$&$	0.199   \pm	0.007 	$&$	19 	\pm	1 	$&$	-2.76   \pm	0.04 	$&$	-5.28 	\pm	0.11 	$&21	& 2\\

W 51 Main/South		&$	49.48 	$&$	-0.39 	$&$	0.185 	\pm	0.010 	$&$	58 	\pm	4 	$&$	-2.64 	\pm	0.16 	$&$	-5.11 	\pm	0.16 	$&32	&\\

IRAS 19213+1723		&$	52.10 	$&$	1.04 	$&$	0.251 	\pm	0.036 	$&$	42 	\pm	3 	$&$	-2.53 	\pm	0.05 	$&$	-6.07 	\pm	0.06 	$&23	&\\

G 59.78+0.06		&$	59.78 	$&$	0.06 	$&$	0.463 	\pm	0.020 	$&$	27 	\pm	3 	$&$	-1.65 	\pm	0.03 	$&$	-5.12 	\pm	0.08 	$&37	&\\

ON1		        &$	69.54 	$&$	-0.98 	$&$	0.404 	\pm	0.017 	$&$	12 	\pm	1 	$&$	-3.10 	\pm	0.18 	$&$	-4.70 	\pm	0.24 	$&20	&\\

G 75.30+1.32		&$	75.30 	$&$	1.32 	$&$	0.108 	\pm	0.005 	$&$	-57 	\pm	2 	$&$	-2.37 	\pm	0.09 	$&$	-4.48 	\pm	0.17 	$&29	&\\
ON2N			&$	75.78 	$&$	0.34 	$&$	0.261 	\pm	0.009 	$&$	0 	\pm	1 	$&$	-2.79 	\pm	0.13 	$&$	-4.66 	\pm	0.17 	$&1	&\\

IRAS 20126+4104		&$	78.12 	$&$	3.64 	$&$	0.610   \pm	0.020 	$&$	-4 	\pm	3 	$&$	-10.25  \pm	0.02 	$&$	-2.30 	\pm	0.28 	$&18	& 1\\
AFGL 2591		&$	78.89 	$&$	0.71 	$&$	0.300   \pm	0.010 	$&$	-6 	\pm	3 	$&$	-1.20   \pm	0.32 	$&$	-4.80 	\pm	0.12 	$&26	& 5\\
IRAS 20290+4052		&$	79.74 	$&$	0.99 	$&$	0.737   \pm	0.062 	$&$	-1 	\pm	3 	$&$	-2.84   \pm	0.09 	$&$	-4.14 	\pm	0.54 	$&26	&\\
DR 20			&$	80.86 	$&$	0.38 	$&$	0.687   \pm	0.038 	$&$	-3 	\pm	3 	$&$	-3.29   \pm	0.13 	$&$	-4.83 	\pm	0.26 	$&26	&\\
DR 21			&$	81.75 	$&$	0.59 	$&$	0.666   \pm	0.035 	$&$	-3 	\pm	3 	$&$	-2.84   \pm	0.15 	$&$	-3.80 	\pm	0.22 	$&26	&\\
W 75 N			&$	81.87 	$&$	0.78 	$&$	0.772   \pm	0.042 	$&$	9 	\pm	3 	$&$	-1.97   \pm	0.10 	$&$	-4.16 	\pm	0.15 	$&26	&\\
AFGL 2789		&$	94.60 	$&$	-1.80 	$&$	0.326   \pm	0.031 	$&$	-44 	\pm	3 	$&$	-2.20   \pm	0.08 	$&$	-3.77 	\pm	0.15 	$&23	& 8\\

IRAS 22198+6336		&$	107.29 	$&$	5.64 	$&$	1.309 	\pm	0.047 	$&$	-11 	\pm	3 	$&$	-2.47 	\pm	0.21 	$&$	0.26 	\pm	0.40 	$&9	&\\

L 1206			&$	108.18 	$&$	5.52 	$&$	1.289 	\pm	0.153 	$&$	-12 	\pm	3 	$&$	0.27 	\pm	0.23 	$&$	-1.40 	\pm	1.95 	$&25	&\\

Cep A HW 2		&$	109.87 	$&$	2.11 	$&$	1.430 	\pm	0.080 	$&$	-10 	\pm	5 	$&$	0.50 	\pm	1.10 	$&$	-3.70 	\pm	0.20 	$&17	&\\

NGC 7538		&$	111.54 	$&$	0.78 	$&$	0.378   \pm	0.017 	$&$	-57 	\pm	3 	$&$	-2.45   \pm	0.03 	$&$	-2.44 	\pm	0.06 	$&17	& 6\\
L 1287			&$	121.29 	$&$	0.66 	$&$	1.077 	\pm	0.039 	$&$	-24 	\pm	3 	$&$	-0.86 	\pm	0.11 	$&$	-2.29 	\pm	0.56 	$&25	&\\

IRAS 00420+5530		&$	122.02 	$&$	-7.07 	$&$	0.460 	\pm	0.010 	$&$	-46 	\pm	3 	$&$	-2.52 	\pm	0.05 	$&$	-0.84 	\pm	0.05 	$&16	&\\

NGC 281			&$	123.07 	$&$	-6.31 	$&$	0.355 	\pm	0.030 	$&$	-31 	\pm	3 	$&$	-2.63 	\pm	0.05 	$&$	-1.86 	\pm	0.08 	$&30	&\\

W 3 OH			&$	133.95 	$&$	1.06 	$&$	0.512 	\pm	0.010 	$&$	-45 	\pm	3 	$&$	-1.20 	\pm	0.20 	$&$	-0.15 	\pm	0.20 	$&36	&\\

S Per			&$	134.62 	$&$	-2.20 	$&$	0.413 	\pm	0.017 	$&$	-39 	\pm	3 	$&$	-0.49 	\pm	0.23 	$&$	-1.19 	\pm	0.20 	$&2	&\\

WB 89-437		&$	135.28 	$&$	2.80 	$&$	0.167 	\pm	0.006 	$&$	-72 	\pm	2 	$&$	-1.22 	\pm	0.30 	$&$	0.46 	\pm	0.36 	$&7	&\\

L 1448 C		&$	157.57 	$&$	-21.94 	$&$	4.310 	\pm	0.330 	$&$	5 	\pm	3 	$&$	21.90 	\pm	0.70 	$&$	-23.10 	\pm	3.30 	$&10	&\\

NGC 1333		&$	158.29 	$&$	-20.50 	$&$	4.250 	\pm	0.320 	$&$	8 	\pm	3 	$&$	14.25 	\pm	2.58 	$&$	-8.95 	\pm	0.74 	$&8	&\\

Hubble 4 (Taurus)	&$	168.84 	$&$	-15.52 	$&$	7.530 	\pm	0.030 	$&$	15 	\pm	2 	$&$	4.30 	\pm	0.05 	$&$	-28.90 	\pm	0.30 	$&34	&\\
IRAS 05168+3634		&$	170.66 	$&$	-0.25 	$&$	0.532 	\pm	0.053 	$&$	-16 	\pm	2 	$&$	0.23 	\pm	1.07 	$&$	-3.14 	\pm	0.28 	$&27	&\\
HP-Tau/G2 (Taurus)	&$	175.73 	$&$	-16.24 	$&$	6.200   \pm	0.030 	$&$	18 	\pm	2 	$&$	13.85   \pm	0.03 	$&$	-15.40 	\pm	0.20 	$&35	& \\
T-Tau/Sb (Taurus)	&$	176.23 	$&$	-20.89 	$&$	6.820   \pm	0.030 	$&$	19 	\pm	1 	$&$	4.02    \pm	0.03 	$&$	-1.18 	\pm	0.05 	$&15	& 4\\

IRAS 06061+2151		&$	188.79 	$&$	1.03 	$&$	0.496     \pm	0.031 	$&$	-2 	\pm	1 	$&$	-0.10   \pm	0.10 	$&$	-3.91 	\pm	0.07 	$&22	& 7\\

S 252/IRAS 06058+2138	&$	188.95 	$&$	0.89 	$&$	0.476 	\pm	0.006 	$&$	11 	\pm	3 	$&$	0.02 	\pm	0.01 	$&$	-2.02 	\pm	0.04 	$&24	&\\

G 192.16$-$3.84		&$	192.16 	$&$	-3.84 	$&$	0.660 	\pm	0.040 	$&$	6 	\pm	3 	$&$	0.84 	\pm	0.16 	$&$	-2.22 	\pm	0.23 	$&33	&\\

S 255			&$	192.60 	$&$	-0.05 	$&$	0.628 	\pm	0.027 	$&$	5 	\pm	1 	$&$	-0.14 	\pm	0.54 	$&$	-0.84 	\pm	1.76 	$&25	&\\

S 269			&$	196.45 	$&$	-1.68 	$&$	0.189 	\pm	0.008 	$&$	20 	\pm	5 	$&$	-0.42 	\pm	0.01 	$&$	-0.12 	\pm	0.04 	$&11	&\\

Orion KL 		&$	209.01 	$&$	-19.38 	$&$	2.390 	\pm	0.030 	$&$	10 	\pm	5 	$&$	9.56 	\pm	0.10 	$&$	-3.83 	\pm	0.15 	$&13	&\\

G 232.6+1.0		&$	232.62 	$&$	1.00 	$&$	0.596 	\pm	0.035 	$&$	23 	\pm	3 	$&$	-2.17 	\pm	0.06 	$&$	2.09 	\pm	0.46 	$&24	&\\

VY CMa			&$	239.35 	$&$	-5.06 	$&$	0.880 	\pm	0.080 	$&$	18 	\pm	3 	$&$	-2.09 	\pm	0.16 	$&$	1.02 	\pm	0.61 	$&5	&\\

S1 (Ophiuchus)		&$	353.10 	$&$	16.89 	$&$	8.550 	\pm	0.500 	$&$	3 	\pm	3 	$&$	-3.88 	\pm	0.87 	$&$	-31.55 	\pm	0.69 	$&15	&\\
$\rho$ Oph East	&$	353.94 	$&$	15.84 	$&$	5.600 	\pm	1.500 	$&$	3 	\pm	3 	$&$    -20.60 	\pm	0.70 	$&$	-32.40 	\pm	2.00 	$&12	&\\

\hline

\\

\multicolumn{8}{@{}l@{}} {\hbox to 0pt{\parbox{170mm}{\footnotesize

(1) \cite{And11} (2) \cite{Asa10} (3) \cite{Bar08} (4) \cite{Bru09} (5) \cite{Cho08} 

(6) \cite{Dzi10} (7) \cite{Hac09} (8) \cite{Hir08a} (9) \cite{Hir08b} (10) \cite{Hir11} 

(11) \cite{Hon07} (12) \cite{Ima07} (13) \cite{Kim08} (14) \cite{Loi07} (15) \cite{Loi08} 

(16) \cite{Moe09} (17) \cite{Mos09} (18) \cite{Mos11} (19) \cite{Mot11} (20) \cite{Nag11a} 

(21) \cite{Nag11b} (22) \cite{Nii11} (23) \cite{Oh10} (24) \cite{Rei09a} (25) \cite{Ryg10} 
(26) \cite{Ryg12} (27) \cite{Sak12} (28) \cite{San09} (29) \cite{San12} (30) \cite{Sat08} 
(31) \cite{Sat10a} (32) \cite{Sat10b} (33) \cite{Shi11} (34) \cite{Tor07} (35) \cite{Tor09} 
(36) \cite{Xu06} (37) \cite{Xu09} (38) \cite{Xu11} (39) \cite{Zha09} 

}\hss}}

    \end{tabular}

  \end{center}

\end{table*}